\documentclass[conference]{IEEEtran}
\IEEEoverridecommandlockouts
\usepackage{cite}
\usepackage{amsmath,amssymb,amsfonts}
\usepackage{algorithmic}
\usepackage{graphicx}
\usepackage{textcomp}
\usepackage{xcolor}
\usepackage{bm}
\usepackage{mathptmx}
\allowdisplaybreaks[4]
\DeclareMathAlphabet{\pazocal}{OMS}{zplm}{m}{n}
\let\Phi\varPhi
\graphicspath{{./figures/}}
\def\BibTeX{{\rm B\kern-.05em{\sc i\kern-.025em b}\kern-.08em
    T\kern-.1667em\lower.7ex\hbox{E}\kern-.125emX}}
\begin{document}

\title{Multivariable Grid-Forming Converters with Direct States Control
\thanks{This work is supported by the VELUX FOUNDATIONS under the VILLUM Investigator Grant REPEPS (Award Ref. No.: 00016591).}
}

\author{\IEEEauthorblockN{Meng Chen, Dao Zhou, Frede Blaabjerg}
\IEEEauthorblockA{\textit{AAU Energy} \\
\textit{Aalborg University}\\
Aalborg, Denmark \\
mche@energy.aau.dk, zda@energy.aau.dk, fbl@energy.aau.dk}
}

\maketitle

\begin{abstract}
A multi-input multi-output based grid-forming (MIMO-GFM) converter has been proposed using multivariable feedback control, which has been proven as a superior and robust system using low-order controllers. However, the original MIMO-GFM control is easily affected by the high-frequency components especially for the converter without inner cascaded voltage and current loops and when it is connected into a strong grid. This paper proposes an improved MIMO-GFM control method, where the frequency and internal voltage are chosen as state variables to be controlled directly. In this way, the impact of high-frequency components is eliminated without increasing the complexity of the control system. The $\pazocal{H}_{\infty}$ synthesis is used to tune the parameters to obtain an optimized performance. Experimental results verify the effectiveness of the proposed method.
\end{abstract}

\begin{IEEEkeywords}
multi-input multi-output grid-forming (MIMO-GFM), direct states control, $\pazocal{H}_{\infty}$ synthesis, power converter, loops coupling
\end{IEEEkeywords}

\section{Introduction}
One of the important features on future power system is the integration of the inverter-based resources (IBRs) due to the concerns on the fossil energy and its impact on the environment \cite{Bose2017}. The ability of self-establishment of the frequency and voltage without relying on external power sources, e.g., synchronous generators, is supposed as essential for some of the IIGs in the future, especially for a system with 100\% IIGs \cite{Deng2021}. To this end, the grid-forming control is a promising solution \cite{Rocabert2012,Rosso2021}.

Up to now, several basic grid-forming controls have been widely researched, e.g., droop control \cite{Chen2018,Mohammed2022}, virtual synchronous generator (VSG) control \cite{Wu2016,Chen2019b,Chen2021b}, power synchronization control \cite{Pan2020}, matching control \cite{Arghir2018}, etc. All of them are proposed based on various assumptions of loops decoupling such as AC power and DC voltage loops, active and reactive power loops. In other words, they try to treat the grid-forming converter as several decoupled single-input single-output (SISO) systems \cite{Wu2016,Chen2019b}. These assumptions simplify the analysis but, at the same time, may sacrifice the performance. Therefore, although the aforementioned grid-forming controls can basically achieve the frequency and voltage regulation, they may not be superior and robust to different operation conditions. To improve the performance of the basic grid-forming controls, several kinds of their improved forms have been proposed, which are usually with higher-order controllers and more complicated control structures to deal with each loops \cite{Liu2020,Chen2021}.

Recently, a new perspective from the multi-input multi-output (MIMO) system to construct and design the grid-forming control has been proposed\cite{Huang2020,chen2021generalized}. In this way, the coupling information among different loops can be used to improve the performance with simple control structures. In \cite{chen2021generalized}, the fundamental theory has been studied using a multivariable feedback control in detail, where a control transfer matrix is used to deal with all the AC power and DC voltage loops as a MIMO integrity and unify different kinds of grid-forming controllers. Thereafter, a MIMO based grid-forming (MIMO-GMF) controller is proposed, which can provide a superior and robust performance without increasing the order of the system. Nevertheless, due to the coupling between the AC power and DC voltage loops, the frequency and internal voltage of the MIMO-GFM converter is sensitive to the high-frequency components of the error signals, which is inevitable. This is because the frequency and internal voltage are not direct state variables. The influence will be more obvious when the grid-forming converter is connected to a strong grid without cascaded voltage and current loops. In practice, pre-filters for decreasing the high-frequency disturbances are usually preferable \cite{Tayyebi2020}, which increase the complexity of the system and may influence the theoretical analysis. This paper extends the work in \cite{chen2021generalized} by proposing a new form of control transfer matrix. The coupling information can still be used to provide a superior and robust performance. Meanwhile, the frequency and internal voltage are chosen as state variables to be controlled. As a result, the impact of the high-frequency components on the frequency and internal voltage is suppressed without increasing the complexity of the control system.

The rest of this paper is organized as follows. In Section II, the preliminaries on MIMO-GFM control is introduced, where its problem is also analyzed in this section. The proposed control and design are given in Section III. In Section IV, experimental results are presented, and finally, conclusions are drawn in Section V.

\section{Preliminary of MIMO-GFM Converters and Problem Definition}

\subsection{Basic Principle of MIMO-GFM Converters}

\begin{figure}[!t]
\centering
\includegraphics[width=\columnwidth]{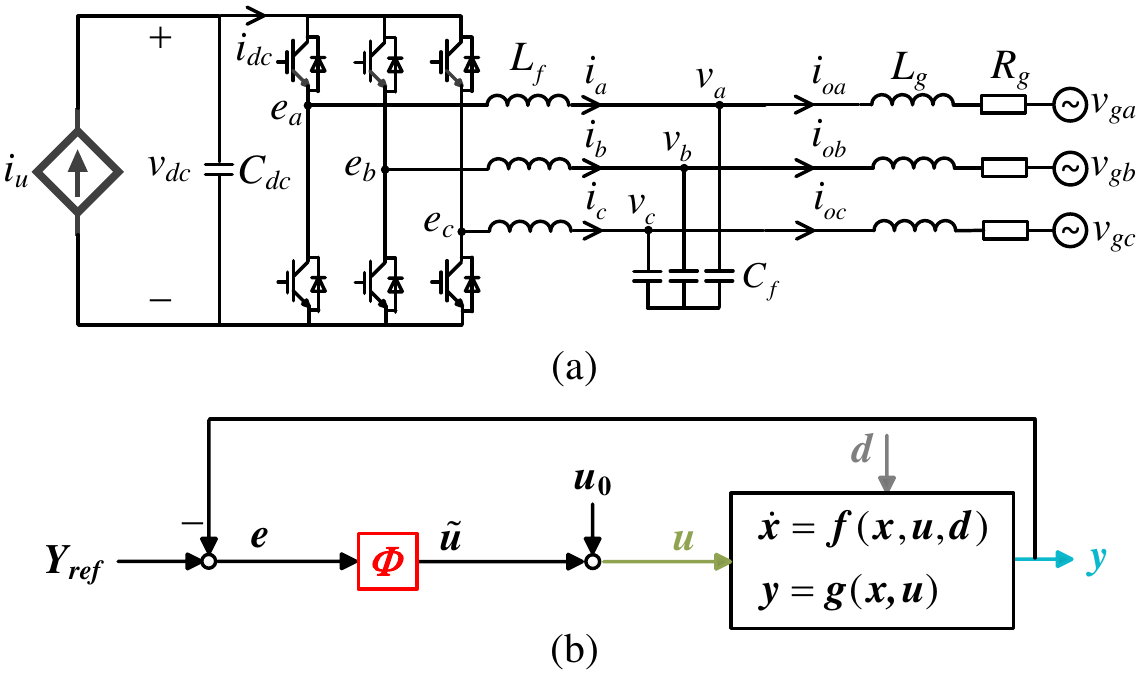}
\caption{Grid-forming converter. (a) Topology. (b) Equivalent closed-loop multivariable feedback control configuration.}
\label{VSC}
\end{figure}

Fig. \ref{VSC} shows the studied topology of the grid-forming converter and its equivalent MIMO closed-loop feedback control configuration. A typical three-phase voltage source inverter is connected to the power grid via an LC filter and a grid line, where $L_f$ and $C_f$ are the inductance and capacitance of the LC filter, $L_g$ and $R_g$ are the equivalent inductance and resistance of the grid line. If an LCL filter is used, the grid-side inductance can be considered into $L_g$. The DC source is represented by a controlled-current source $i_u$ paralleled with a DC capacitor, where the capacitance is $C_{dc}$.

Then the state-space model of the topology can be built in the $d$-$q$ frame and it is represented in a compact form in Fig. \ref{VSC}(b) as
\begin{align}
&\dot{\bm{x}}=\bm{f}(\bm{x},\bm{u},\bm{d})\\
&\bm{y}=\bm{g}(\bm{x},\bm{u})
\end{align}
and the vectors are defined as
\begin{align}
&\bm x=\left[
	\begin{matrix}
	i_d&i_q&v_d&v_q&i_{od}&i_{oq}&\delta&v_{dc}
	\end{matrix}
\right]^T\\
&\bm u=\left[
	\begin{matrix}
	i_u&\omega_u&E_u
	\end{matrix}
\right]^T\\
&\bm y=\left[
	\begin{matrix}
	v_{dc}&p&\omega_u&q&V
	\end{matrix}
\right]^T\\
&\bm d=\left[
	\begin{matrix}
	\omega_g&V_g
	\end{matrix}
\right]^T,
\end{align}
where $i_{dq}$ are the currents of the filter inductor, $v_{dq}$ are voltages of the filter capacitor, $i_{odq}$ are the output currents, $\delta$ is the angle difference between the grid-forming converter internal voltage and the grid voltage, $v_{dc}$ is the DC voltage, $\omega_u$ and $E_u$ are the frequency and internal voltage to be obtained by the grid-forming controller, $p$ and $q$ are the output active and reactive powers, respectively, $V$ is the magnitude of the capacitor voltage, $\omega_g$ and $V_g$ are the frequency and magnitude of the grid voltage. Moreover, the detailed mathematical model of the system can be given as
\begin{align}
&\dot i_d=\frac{\omega_b}{L_f}E_u-\frac{\omega_b}{L_f}v_d+\omega_b\omega_ui_q\\
&\dot i_q=-\frac{\omega_b}{L_f}v_q-\omega_b\omega_ui_d\\
&\dot v_d=\frac{\omega_b}{C_f}i_d-\frac{\omega_b}{C_f}i_{od}+\omega_b\omega_uv_q\\
&\dot v_q=\frac{\omega_b}{C_f}i_q-\frac{\omega_b}{C_f}i_{oq}-\omega_b\omega_uv_d\\
&\dot i_{od}=\frac{\omega_b}{L_g}v_d-\frac{\omega_b}{L_g}V_g\cos\delta-\frac{\omega_bR_g}{L_g}i_{od}+\omega_b\omega_ui_{oq}\\
&\dot i_{oq}=\frac{\omega_b}{L_g}v_q+\frac{\omega_b}{L_g}V_g\sin\delta-\frac{\omega_bR_g}{L_g}i_{oq}-\omega_b\omega_ui_{od}\\
&\dot\delta=\omega_b\omega_u-\omega_b\omega_g\\
&\dot v_{dc}=\frac{\omega_b}{C_{dc}}i_u-\frac{\omega_bE_ui_d}{C_{dc}v_{dc}}\\
&p=v_di_{od}+v_qi_{oq}\\
&q=-v_di_{oq}+v_qi_{od}\\
&V=\sqrt{v^2_d+v^2_q}
\end{align}

In Fig. \ref{VSC}(b), $\bm{\Phi}(s)=(\phi_{ij})_{3\times 5}$ is a control transfer matrix, which copes with all the loops as a MIMO integrity rather than decoupled SISO systems. In \cite{chen2021generalized}, $\bm{\Phi}$ is designed as follows for the MIMO-GFM converter
\begin{align}
\label{phi_mimo}
\bm\Phi_{MIMO}=\left[
	\begin{matrix}
	k_{pdc}+\frac{k_{idc}}{s}&k_{12}&0&k_{14}&k_{15}\\
	k_{21}&D_pk_{22}/(s+k_{22})&0&k_{24}&\frac{k_{24}}{D_q}\\
	k_{31}&k_{32}&0&k_{34}/s&\frac{k_{34}/D_q}{s}
	\end{matrix}
\right]
\end{align}
where $D_p$ and $D_q$ are the droop coefficients for active and reactive power controls, respectively. As shown, the MIMO-GFM controller keeps all the favorable features of the basic controllers, i.e., inertia and droop characteristics. Meanwhile, it uses several simple proportional gains to deal with the coupling terms, which has been proved as an effective way to provide a superior and robust performance without increasing the order of the system.

\subsection{Impact of High-Frequency Components}

To better illustrate the problem of the MIMO-GFM controller, we will first consider a basic VSG control. The control transfer matrix $\Phi$ can be derived by setting all the coupling gains of (\ref{phi_mimo}) as zero
\begin{align}
\bm\Phi_{VSG}=\left[
	\begin{matrix}
	k_{pdc}+k_{idc}/s&0&0&0&0\\
	0&D_pk_{22}/(s+k_{22})&0&0&0\\
	0&0&0&k_{34}/s&\frac{k_{34}/D_q}{s}
	\end{matrix}
\right]
\end{align}
based on which the state differential equations of $\Phi_{VSG}$ can be expressed as
\begin{align}
\label{xphivsg1}
&\dot{x}_{\phi VSG1}=k_{idc}e_1\\
&\dot{x}_{\phi VSG2}=-k_{22}x_{\phi VSG2}+D_pk_{22}e_2\\
\label{xphivsg3}
&\dot{x}_{\phi VSG3}=k_{34}e_4+\frac{k_{34}}{D_q}e_5
\end{align}
and the state variables are defined as
\begin{align}
\label{xphivsg}
\bm{x_{\phi VSG}}=\left[
	\begin{matrix}
	\tilde{i}_u-k_{pdc}e_1&\tilde\omega_u&\tilde{E}_u
	\end{matrix}
\right]^T
\end{align}

Similarly, the state differential equations of $\Phi_{MIMO}$ in (\ref{phi_mimo}) can be expressed as
\begin{align}
\label{xphimimo1}
&\dot{x}_{\phi MIMO1}=k_{idc}e_1\\
&\dot{x}_{\phi MIMO2}=-k_{22}x_{\phi MIMO2}+D_pk_{22}e_2\\
\label{xphimimo3}
&\dot{x}_{\phi MIMO3}=k_{34}e_4+\frac{k_{34}}{D_q}e_5
\end{align}
and the state variables are defined as
\begin{align}
\label{xphimimo}
\bm{x_{\phi MIMO}}=\left[
	\begin{matrix}
	\tilde{i}_u-k_{pdc}e_1-k_{12}e_2-k_{14}e_4-k_{15}e_5\\
	\tilde\omega_u-k_{21}e_1-k_{24}e_4-\frac{k_{24}}{D_q}e_5\\
	\tilde{E}_u-k_{31}e_1-k_{32}e_2
	\end{matrix}
\right]
\end{align}

The following conclusions can be summarized.
\begin{enumerate}
	\item From (\ref{xphivsg1})-(\ref{xphivsg}), the control of the basic VSG is decoupled, i.e., $\tilde i_u$ is controlled by $e_1$, $\tilde\omega_u$ is controlled by $e_2$, $\tilde{E}_u$ is controlled by $e_4$ and $e_5$.
	\item Comparing (\ref{xphivsg1})-(\ref{xphivsg3}) with (\ref{xphimimo1})-(\ref{xphimimo3}), it is observed that the state differential equations of the VSG and MIMO-GFM controls have the same structure.
	\item Comparing (\ref{xphivsg}) with (\ref{xphimimo}), the state variables between the VSG and MIMO-GFM controls are quite differnt.
\end{enumerate}

Therefore, it is clear that the principle of the MIMO-GFM control is to change the state variables by adding the coupling terms but not to change the form of the state differential equations compared with the basic VSG control. These coupling terms are expected to improve the performance because they provide a multi-degree-of-freedom control and the useful coupling information among various loops can be considered as well.

\begin{figure}[!t]
\centering
\includegraphics[width=\columnwidth]{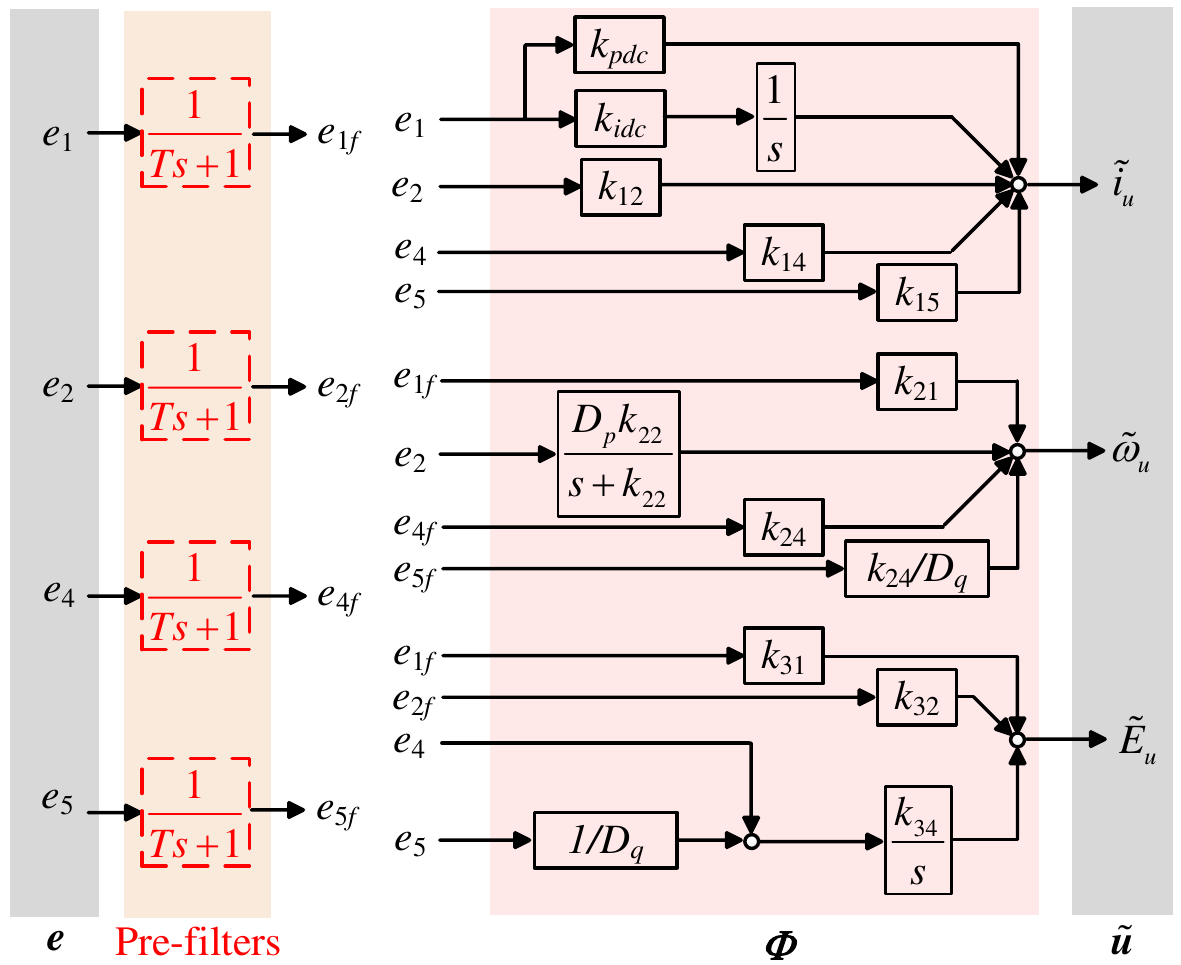}
\caption{Practical control block diagram of MIMO-GFM converter with pre-filters.}
\label{MIMO_GFM}
\end{figure}

Nevertheless, due to the changes of the state variables, the frequency $\tilde\omega_u$ and internal voltage $\tilde{E}_u$ are not the state variables to be directly controlled anymore for the MIMO-GFM converter. Instead, they will be directly influenced by the errors $\bm e$ as shown in (\ref{xphimimo}), especially the high-frequency components, which are inevitable. On one hand, there are always high-frequency components in the steady-state powers  and DC voltage. On the other hand, steps of the references $\bm Y_{ref}$ will also introduce the high-frequency components into the errors. Therefore, in practice, pre-filters are usually used before the coupling terms in order to suppress the influence of the high-frequency components as shown in Fig. \ref{MIMO_GFM}. However, these pre-filters increase the order and complexity of the system and may highly influence the stability.

\section{Proposed Direct States Control}

\subsection{Principle of Proposed Direct States Control}

According to the aforementioned analysis, the problem of the MIMO-GFM control is due to the fact that the added coupling terms change the state variables compared with the basic VSG control. Motivated from this point, an improved MIMO-GFM control is proposed, where the state differential equations of the control transfer matrix is designed as
\begin{align}
\label{xphiproposed1}
&\dot{x}_{\phi 1}=-k_{12}x_{\phi 2}+k_{idc}e_1+D_pk_{12}e_2+k_{14}e_4+\frac{k_{14}}{D_q}e_5\\
&\dot{x}_{\phi 2}=-k_{22}x_{\phi 2}+k_{21}e_1+D_pk_{22}e_2+k_{24}e_4+\frac{k_{24}}{D_q}e_5\\
\label{xphiproposed3}
&\dot{x}_{\phi 3}=-k_{32}x_{\phi 2}+k_{31}e_1+D_pk_{32}e_2+k_{34}e_4+\frac{k_{34}}{D_q}e_5
\end{align}
and the state variables are defined as
\begin{align}
\label{xphiproposed}
\bm{x_{\phi}}=\left[
	\begin{matrix}
	\tilde{i}_u-k_{pdc}e_1&\tilde\omega_u&\tilde{E}_u
	\end{matrix}
\right]^T
\end{align}

The following conclusions about the proposed control can be summarized.
\begin{enumerate}
	\item Comparing (\ref{xphiproposed}) with (\ref{xphivsg}), the proposed control chooses the same state variables as the basic VSG control. Especially, the frequency and internal voltage are still directly controlled state variables. 
	\item Comparing (\ref{xphiproposed1})-(\ref{xphiproposed3}) with (\ref{xphivsg1})-(\ref{xphivsg3}) and (\ref{xphimimo1})-(\ref{xphimimo3}), it is observed that the state differential equations of the proposed control has a different structure, which is taking the coupling terms into consideration.
\end{enumerate}

In this way, the proposed control can not only suppress the influence of the high-frequency components but also improve the performance using the coupling terms. The corresponding control transfer matrix of the proposed method can be derived as
\begin{align}
\bm\Phi(s)=(\phi_{ij})_{3\times 5}
\end{align}
where
\begin{align}
&\phi_{11}=\frac{k_{pdc}s^2+(k_{pdc}k_{22}+k_{idc})s+k_{idc}k_{22}-k_{12}k_{21}}{s^2+k_{22}s}\\
&\phi_{12}=\frac{D_pk_{12}}{s+k_{22}},~\phi_{14}=\frac{k_{14}s+k_{14}k_{22}-k_{12}k_{24}}{s^2+k_{22}s}\\
&\phi_{15}=\frac{(k_{14}/D_q)s+k_{14}k_{22}/D_q-k_{12}k_{24}/D_q}{s^2+k_{22}s}\\
&\phi_{21}=\frac{k_{21}}{s+k_{22}},~\phi_{22}=\frac{D_pk_{22}}{s+k_{22}}\\
&\phi_{24}=\frac{k_{24}}{s+k_{22}},~\phi_{25}=\frac{k_{24}/D_q}{s+k_{22}}\\
&\phi_{31}=\frac{k_{31}s+k_{22}k_{31}-k_{21}k_{32}}{s^2+k_{22}s},~\phi_{32}=\frac{D_pk_{32}}{s+k_{22}}\\
&\phi_{34}=\frac{k_{34}s+k_{22}k_{34}-k_{24}k_{32}}{s^2+k_{22}s}\\
&\phi_{35}=\frac{(k_{34}/D_q)s+k_{22}k_{34}/D_q-k_{24}k_{32}/D_q}{s^2+k_{22}s}\\
&\phi_{13}=\phi_{23}=\phi_{33}=0
\end{align}

\begin{figure}[!t]
\centering
\includegraphics[width=\columnwidth]{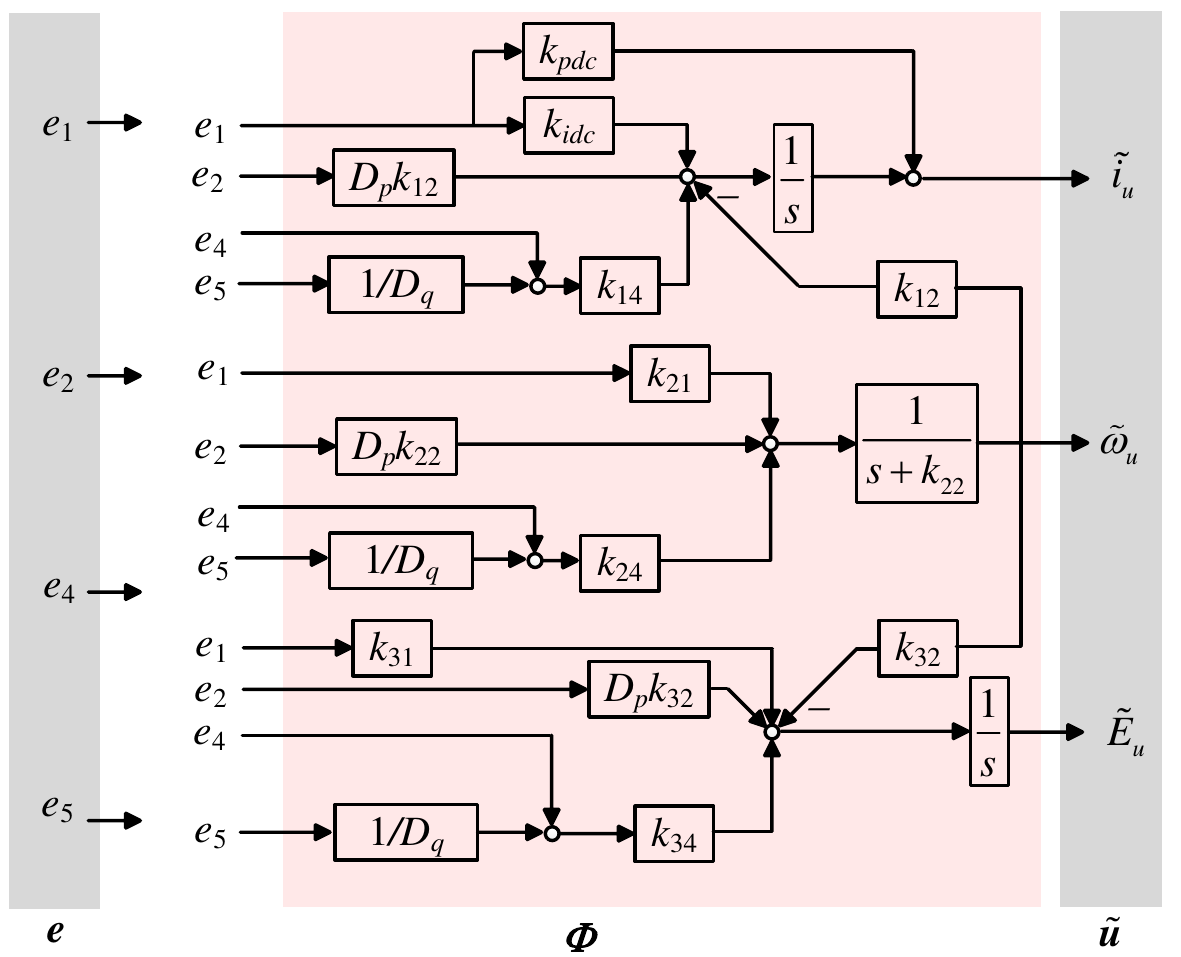}
\caption{Block diagram of proposed control transfer matrix $\Phi$.}
\label{proposed}
\end{figure}

According to the aforementioned analysis, the block diagram of the proposed control transfer matrix is as shown in Fig. \ref{proposed}. It is worth mentioning that although the above elements of $\phi_{ij}$ seem to have complicated forms, according to (\ref{xphiproposed1})-(\ref{xphiproposed}), the order of the system is not increased and the control structure is straightforward as well, which is because $\phi_{ij}$ actually share many common parts as shown in Fig. \ref{proposed}. As observed, it still only uses simple proportional controllers to include the coupling terms. 

\subsection{Parameters Design based on $\pazocal{H}_{\infty}$ Optimization}

\begin{figure}[!t]
\centering
\includegraphics[width=\columnwidth]{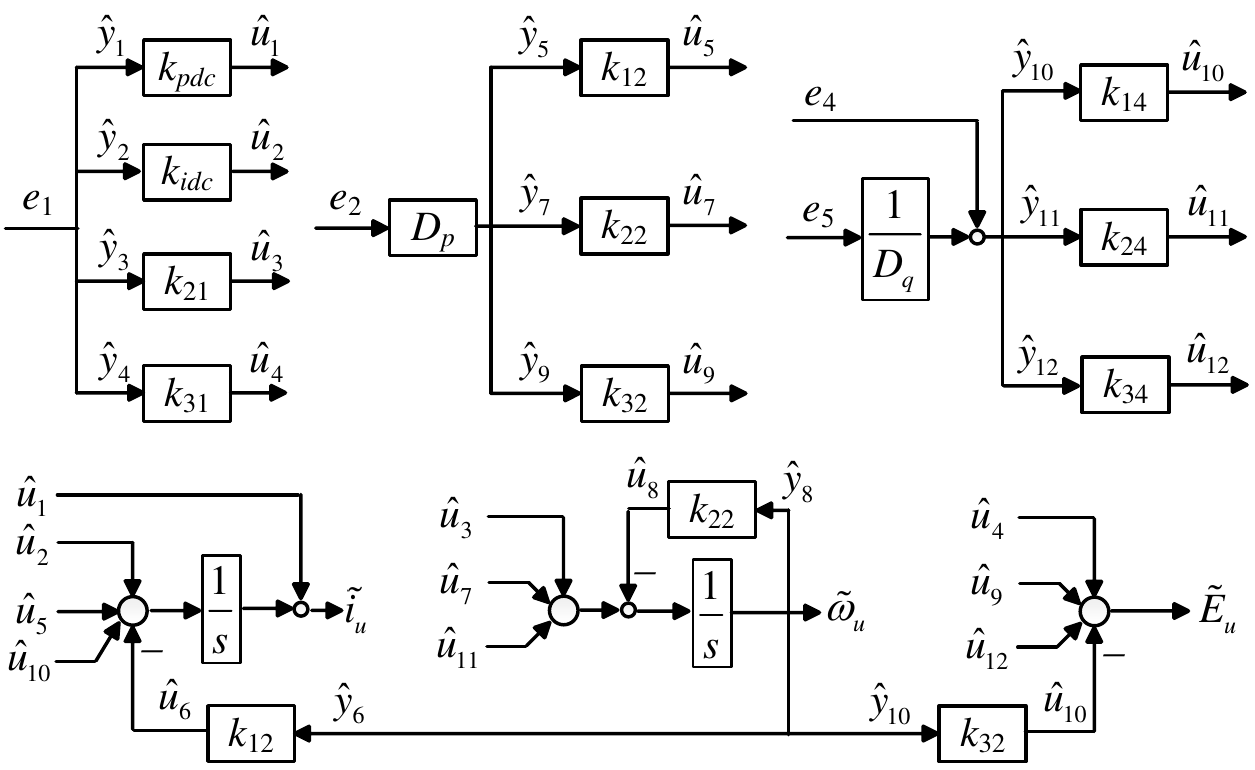}
\caption{Equivalent block diagram of proposed control transfer matrix for the formulation of $\pazocal{H}_{\infty}$ synthesis.}
\label{h_infinity_formulation}
\end{figure}

\begin{figure}[!t]
	\centering
	\includegraphics[width=0.4\columnwidth]{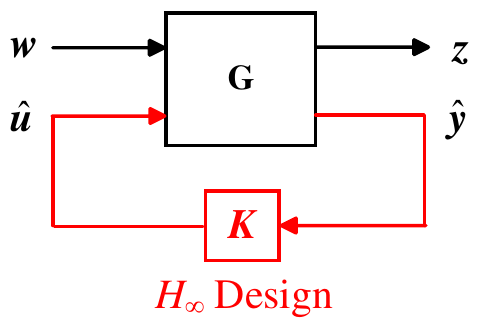}
	\caption{Block diagram of grid-forming converter in linear fractional transformation.}
	\label{LFT}
\end{figure}

To make a fair comparison with the original MIMO-GFM control, this paper also uses the $\pazocal{H}_{\infty}$ synthesis to tune the parameters to obtain an optimal performance \cite{chen2021generalized}. Therefore, the block diagram of the proposed control transfer matrix in Fig. \ref{proposed} is equivalently changed to Fig. \ref{h_infinity_formulation} by defining two intermediate vectors $\hat{\bm u}$ and $\hat{\bm y}$, which have the following relationship
\begin{align}
\label{u}
\hat{\bm u}=&diag(k_{pdc},k_{idc},k_{21},k_{31},k_{12}\bm{I_2},k_{22}\bm{I_2},k_{32}\bm{I_2},k_{14},k_{24},k_{34})\hat{\bm y}\notag\\
=&\bm K\hat{\bm y}
\end{align}
where $\bm K$ is a gain vector only containing all the parameters to be tuned. Thereafter, the standard form of linear fractional transformation for $\pazocal{H}_{\infty}$ optimization can be obtained as shown in Fig. \ref{LFT}, where the grid-forming converter in Fig. \ref{VSC}(b) is collapsed into \textbf{G} (except for $\bm K$). The disturbance inputs and evaluation outputs for the $\pazocal{H}_{\infty}$ synthesis are defined as
\begin{align}
\label{w}
&\bm w=\left[
	\begin{matrix}
	P_{ref}&\omega_g
	\end{matrix}
\right]^T\\
\label{z}
&\bm z= \left[
	\begin{matrix}
	P_{ref}-p&p&\omega_u&q+V/D_q
	\end{matrix}
\right]^T
\end{align}
where the transfer functions $T_{ij}(s)$ from $w_j$ to $z_i$ are limited by the following chosen weighting functions $W_{ij}(s)$
\begin{align}
\label{w11}
&W_{11}(s)=\frac{s+4}{s+0.0004}\\
&W_{21}(s)=\left({\frac{1.447\times 10^{-3}s+1}{1.447\times 10^{-5}s+1}}\right)^2\\
&W_{22}(s)=\frac{1}{100}\times\frac{1.447\times 10^{-3}s+1}{1.447\times 10^{-5}s+1}\\
&W_{31}(s)=\frac{1}{0.015}\times\frac{s}{1.447\times 10^{-5}s+1}\\
&W_{32}(s)=\frac{1.447\times 10^{-3}s+1}{1.447\times 10^{-5}s+1}\\
&W_{41}=\frac{s+60}{s+0.006}
\end{align}
and the considerations of choosing the weighting functions can be found in \cite{chen2021generalized}. Finally, the parameters can be derived by solving the following $\pazocal{H}_{\infty}$ optimization problem
\begin{align}
\min_{\bm K}||diag(W_{ij}(s)T_{ij}(s))||_\infty
\end{align}

\begin{figure}[!t]
	\centering
	\includegraphics[width=\columnwidth]{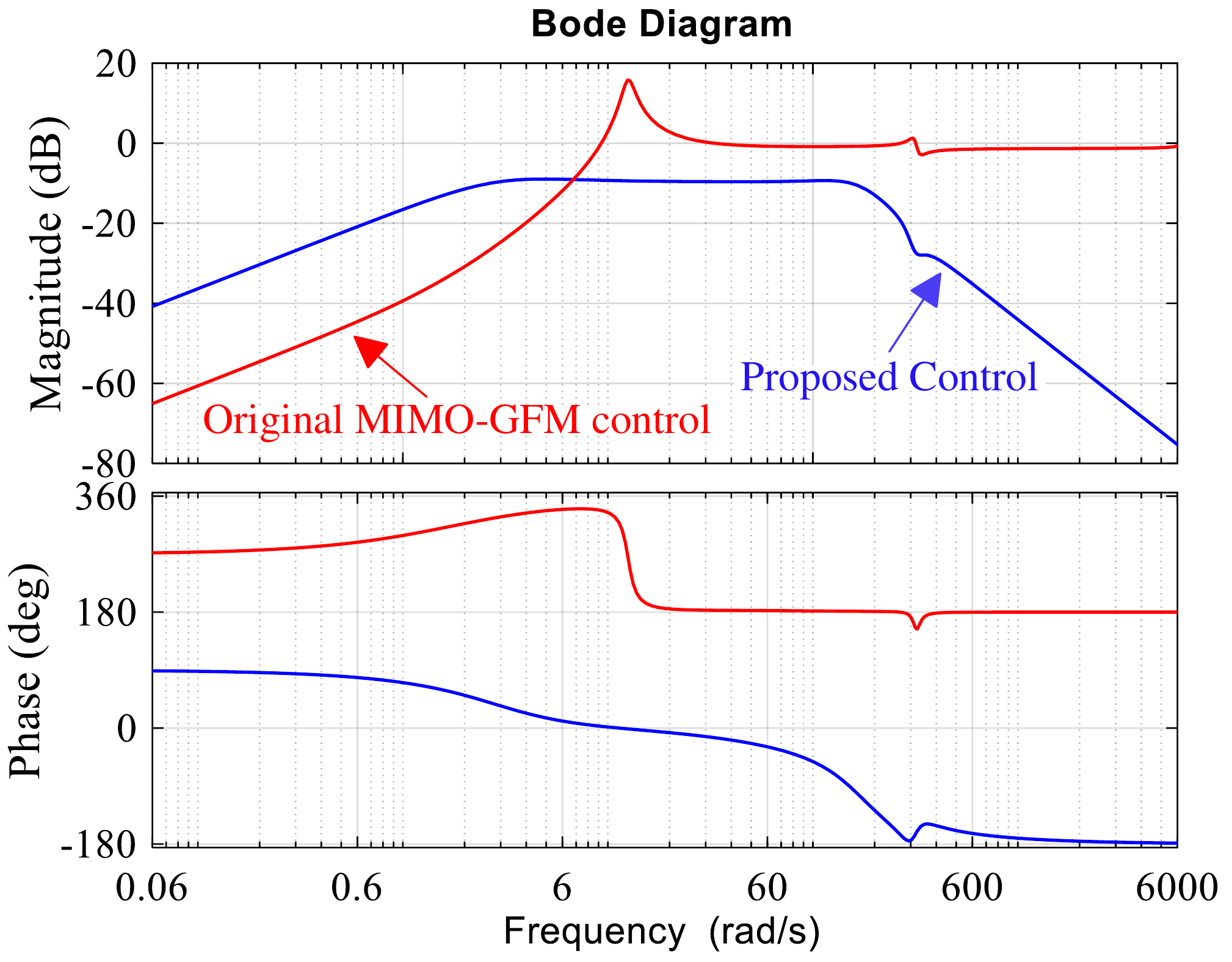}
	\caption{Comparison of bode diagram between original MIMO-GFM control and proposed control without any pre-filter.}
	\label{bode}
\end{figure}

According to the aforementioned method with the parameters listed in Table \ref{parameter}, the parameters of the proposed control can be derived as shown in Table \ref{parameter_mimo}. For the following comparison, the parameters of the original MIMO-GFM control are also presented in Table \ref{parameter_mimo}. Fig. \ref{bode} compares the bode diagrams of the closed-loop transfer function from $e_1$ to $\omega_u$. As observed, the original MIMO-GFM control without those pre-filters fails to suppress the components with the frequency over 200 rad/s. In comparison, the proposed control make the log-magnitude curve to decrease with a slope of -40 dB/decade even though we have neither defined an explicit weighting function to limit the corresponding high-frequency components nor added any pre-filter, which verify the advantages of the proposed control method.

\begin{table}[!t]
	\renewcommand{\arraystretch}{1.3}
	\caption{Parameters of Test System}
	\centering
	\label{parameter}
	\resizebox{\columnwidth}{!}{
		\begin{tabular}{c l c}
			\hline\hline \\[-3mm]
			Symbol & Description & Value  \\ \hline
			$\omega_n$  & nominal frequency & $100\pi$ rad/s \\
			$S_n $ & nominal power &  4 kW  \\ 
			$V_n $ & nominal line-to-line RMS voltage & 380 V \\
			$f_{sw}$ & switching frequency & 10 kHz \\
			$\omega_g $ & grid frequency & $100\pi$ rad/s (1 p.u.)  \\
			$V_g $ & grid voltage & 380 V (1 p.u.) \\
			$L_g$  & line inductor & 2 mH (0.0174 p.u.)\\
			$R_g$ & filter resistor & 0.06 $\Omega$ (0.0017 p.u.) \\ 
			$C_f$ & filter capacitor & 20 $\mu$F (0.2268 p.u.)\\   
			$L_f$ & filter inductor & 2 mH (0.0174 p.u.)\\
			$R_f$ & filter resistor & 0.06 $\Omega$ (0.0017 p.u.) \\ 
			$C_{dc}$ & DC capacitor & 500 $\mu$F \\ 
			$D_p$ & droop coefficient of $P$-$f$ regulation & 0.01 p.u. \\ 
			$D_q$ &droop coefficient of $Q$-$V$ regulation & 0.05 p.u.\\
			$P_{ref}$ & Active power reference & 0.5 p.u.\\
			$Q_{ref}$ & Reactive power reference & 0 p.u.\\
			$V_{ref}$ & Voltage magnitude reference & 1 p.u.\\
			$V_{dcref}$ &DC voltage reference & 700 V\\[1.4ex]
			\hline\hline
		\end{tabular}
	}
\end{table}

\begin{table}[!t]
	\renewcommand{\arraystretch}{1.3}
	\caption{Parameters of Control Transfer Matrix for Original MIMO-GFM and Proposed Controllers}
	\centering
	\label{parameter_mimo}
	\resizebox{\columnwidth}{!}{
		\begin{tabular}{c c  c}
			\hline\hline \\[-4mm]
			Parameters & Original MIMO-GFM Control & Proposed Control   \\ \hline
			$k_{pdc}$  & 120.224 & 18.8801  \\
			$k_{idc}$  & 265.6217 &   2811.2 \\ 
			$k_{12}$  & -0.0019 &  123.7138\\
			$k_{14}$  & 0.1673 &  4.9404\\
			$k_{15}$  & -0.8274 &   - \\
			$k_{21}$ & -0.8382&-20.1083\\
			$k_{22}$ & 1.7622&0.5532\\
			$k_{24}$ & 0&0.0615\\
			$k_{31}$ & -4.8977 &5.684\\
			$k_{32}$ & 0 &-0.1862\\
			$k_{34}$ & 1.0844& 0.0908\\[0.5ex]
			\hline\hline
		\end{tabular}
	}
\end{table}

\section{Experimental Results}
\begin{figure}[!t]
\centering
\includegraphics[width=\columnwidth]{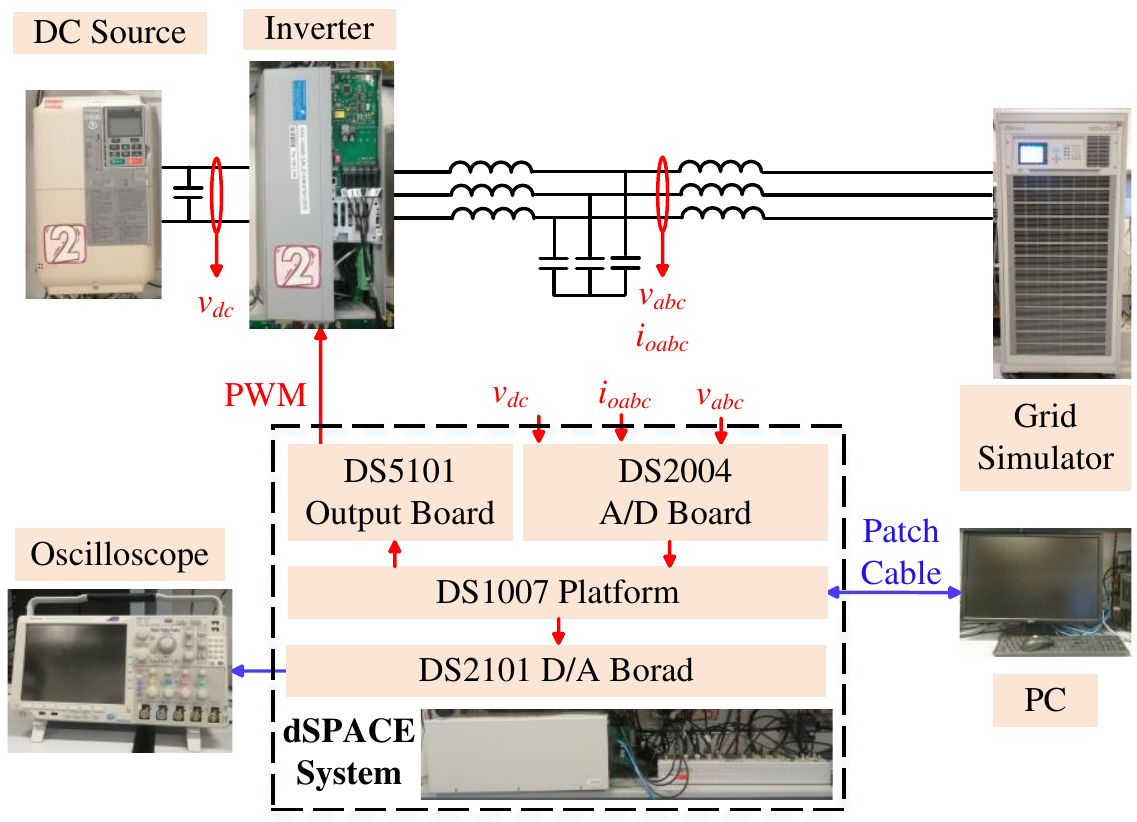}
\caption{Experimental configuration of MIMO-GFM converter.}
\label{setup}
\vspace{-10pt}
\end{figure}

The performance of the proposed control is tested by using the experimental setup shown in Fig. \ref{setup}. The parameters are the same as those in Table \ref{parameter} and Table \ref{parameter_mimo}, which represents a system of a grid-forming converter connected to a strong grid. The structure of the control system is same as Fig. \ref{VSC} without cascaded voltage and current loops. Meanwhile, due to the DC control is fixed in the DC source, only the power loops can be tested. Nevertheless, it does not influence the conclusion.

\begin{figure}[!t]
\centering
\includegraphics[width=\columnwidth]{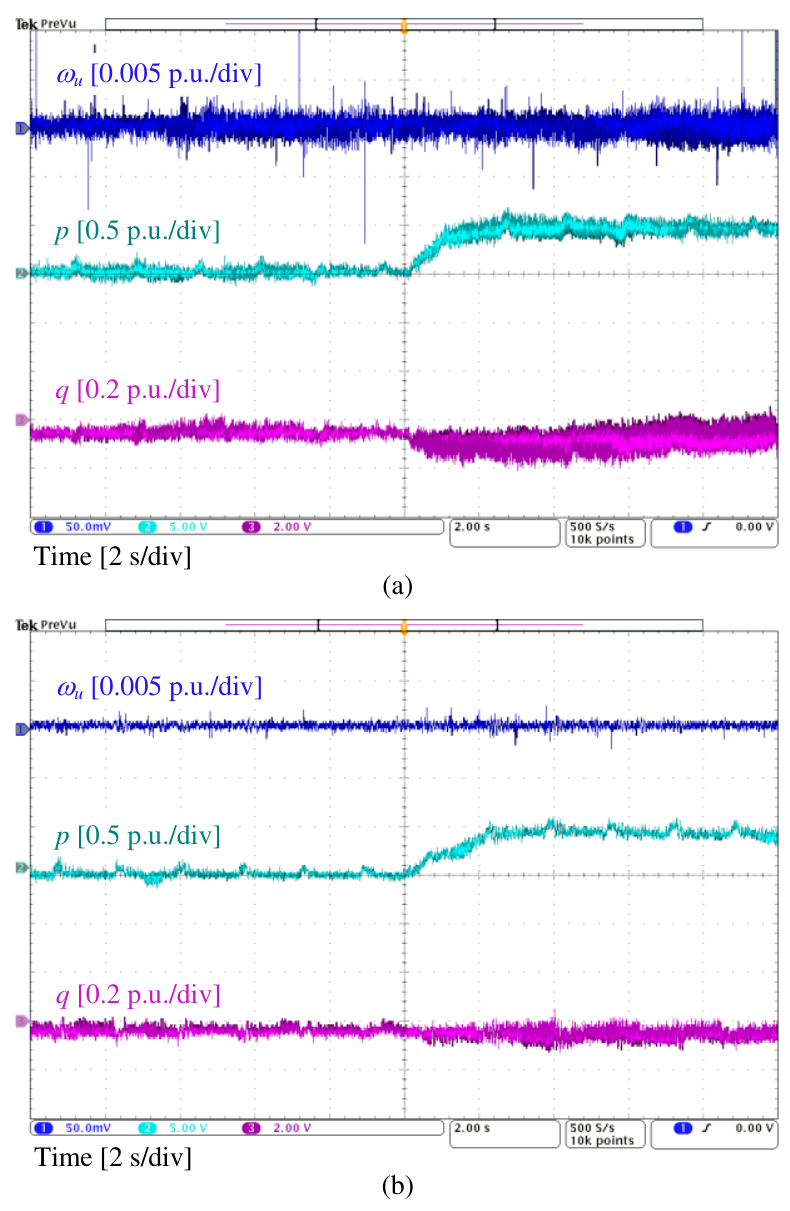}
\caption{Experimental comparison when $P_{ref}$ steps from 0.5 p.u. to 1 p.u. (a) Original MIMO-GFM control. (b) Proposed MIMO-GFM control using direct state control.}
\label{experiment_pref}
\vspace{-10pt}
\end{figure}

Fig. \ref{experiment_pref} compares the experimental results with the original MIMO-GFM control and the proposed control when $P_{ref}$ steps from 0.5 p.u. to 1 p.u. It should be mentioned that no any pre-filter is included in the control system. As shown, although the original MIMO-GFM control can guarantee the stability and good low-frequency dynamics, the focused variables have large high-frequency components especially in the frequency $\omega_u$. Therefore, as in the aforementioned discussion, pre-filters are necessary. In comparison, the proposed control can highly damp the high-frequency components without additional means.

\begin{figure}[!t]
\centering
\includegraphics[width=\columnwidth]{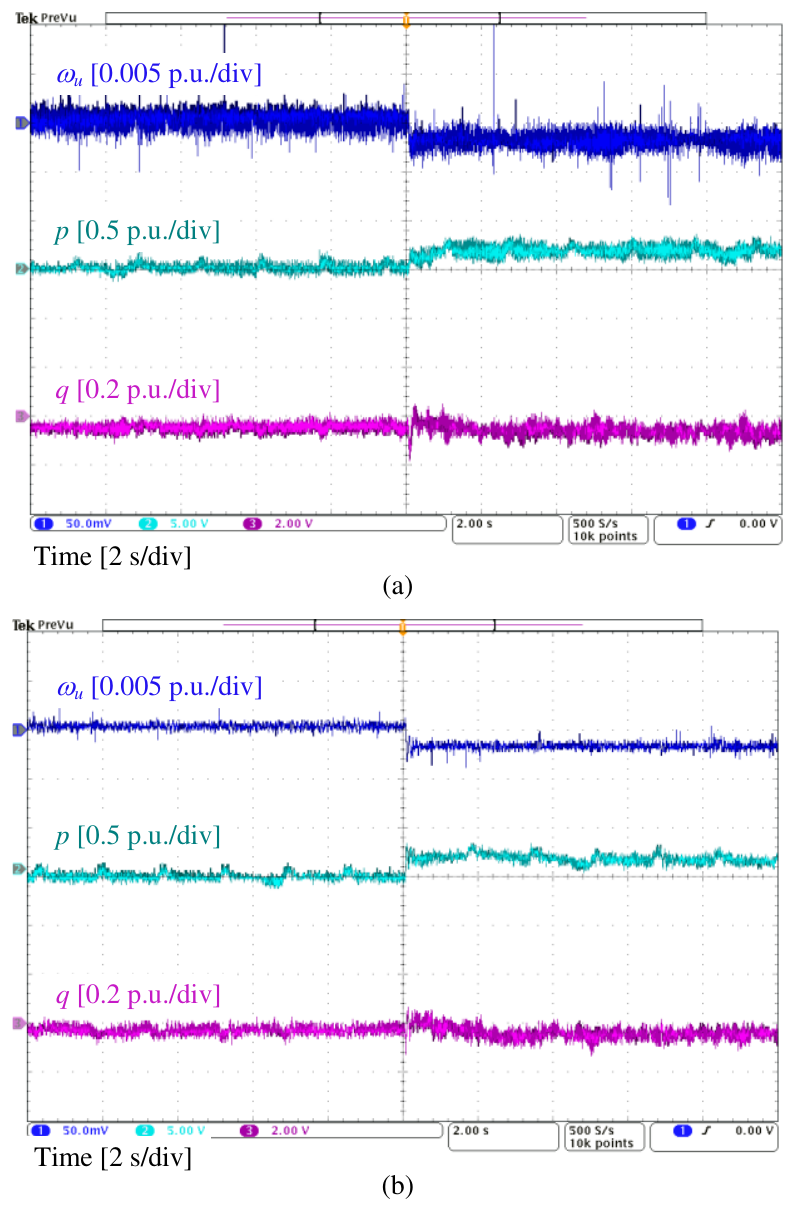}
\caption{Experimental comparison when $\omega_g$ decreases from 50 Hz to 49.9 Hz. (a) Original MIMO-GFM control. (b) Proposed MIMO-GFM control using direct state control}
\label{experiment_wg}
\vspace{-10pt}
\end{figure}

A further comparison when $\omega_g$ decreases from 50 Hz to 49.9 Hz is shown in Fig. \ref{experiment_wg}. Similar to Fig. \ref{experiment_pref}, the proposed control can well damp the high-components, which is obviously observed with the original MIMO-GFM control. The results of Fig. \ref{experiment_pref} and \ref{experiment_wg} are in accordance with the theoretical analysis of Fig. \ref{bode}. 

\section{Conclusion}
This paper proposes a novel control transfer matrix for the MIMO-GFM control. Instead of changing the states, the coupling terms of the proposed method only change the structure of the state differential equations and, at the same time, keep the frequency and voltage as the controlled states. By designing with the same $\pazocal{H}_{\infty}$ synthesis as the original MIMO-GFM control, the proposed method will have improved ability to decrease the impact of the high-frequency components on the system dynamics without increasing the complexity of the control system.
\bibliographystyle{IEEEtran}
\bibliography{IEEEabrv,ECCE}

\end{document}